# Lattice Instability of UTe$_2$ Studied by Ultrasonic Measurements

Keita USHIDA[1], Tatsuya YANAGISAWA[1]*, Ruo HIBINO[1], Masato MATSUDA[1], Hiroyuki HIDAKA[1], Hiroshi AMITSUKA[1], Georg KNEBEL[2], Jacques FLOUQUET[2], and Dai AOKI[3]

[1]*Department of Physics, Hokkaido University, Sapporo 060-0810, Japan*
[2]*Univ. Grenoble Alpes, Grenoble INP, CEA, IRIG-PHELIQS, F-38000 Grenoble, France*
[3]*Institute for Materials Research, Tohoku University, Oarai 311-1313, Japan*

*E-mail: tatsuya@phys.sci.hokudai.ac.jp

The elastic constants of an unconventional superconductor, UTe$_2$, were investigated using ultrasound. In this paper, we report the elastic response of the normal state at temperatures down to 2 K and up to 14 T for $H \parallel b$ at ambient pressure. The transverse ultrasonic mode $C_{55}$, which corresponds to the strain susceptibility of $\varepsilon_{zx}$, shows softening with decreasing temperature, whereas the $C_{44}$ and $C_{33}$ modes do not show such softening. This clear mode dependence strongly suggests that UTe$_2$ has a lattice instability for this specific symmetry.

**KEYWORDS:** UTe$_2$, reentrant superconductivity, spin triplet, elastic constants, lattice instability

## 1. Introduction

Spin-singlet superconductivity is generally fragile against magnetic fields because the Cooper pairs are destroyed by the external or internal magnetic field when the magnetic field exceeds a certain threshold, the so-called Pauli limit. On the other hand, there is a different type of superconducting state in which the Cooper pairs do not cancel out their spins, which is named spin-triplet superconductivity. Possible spin-triplet superconductor candidates have been found among uranium-based ferromagnetic compounds, in which the superconducting phase coexists with the ferromagnetic ordering state. For example, URhGe [1] and UCoGe [2] show superconductivity in the ferromagnetic phase even at ambient pressure, where unique phenomena cannot be explained by the BCS theory, *e.g.,* the upper critical field of the superconducting phase significantly exceeds the Pauli limit. One of the most important features of ferromagnetic superconductors is their magnetic field reentrant superconductivity. The superconductivity of URhGe shows reentrant behavior when the field is applied along the *b*-axis of its orthorhombic structure [3].

Recently, the heavy fermion superconductor UTe$_2$, which has an orthorhombic structure (No. 71, *Immm*, $D_{2h}^{25}$), has attracted considerable attention as a new candidate for the spin-triplet superconductor, and this compound is paramagnetic down to 25 mK. Because of its similarity to the ferromagnetic compounds mentioned above, UTe$_2$ is initially considered to be located near ferromagnetic instability. However, antiferromagnetic fluctuation has been experimentally detected in UTe$_2$ [4,5]. The superconductivity of UTe$_2$ also shows two different types of reentrant behavior under the magnetic field $H$ applied along the $b$-axis and between the $b$-axis and the $c$-axis with an angular-range from ∼ 20° to 40°. In the latter case, the magnetic-field-induced superconducting phase exists at least up to $H \sim 60$ T [6,7].

Elastic properties of UTe$_2$ have been investigated by thermal expansion measurements [8,9]. The linear thermal expansion coefficients for all three crystallographic axes become negative below 30 K. Thermal expansion measurement mainly provides information on volume changes that preserve symmetry. On the other hand, ultrasonic measurement is a powerful experimental technique for investigating the lattice instability in the symmetry breaking mode as strain susceptibility, which will provide us additional information to understand the 5$f$-electronic state of the present system via elastic properties. The elastic constants near the superconductivity phase have been determined by the resonant ultrasound spectroscopy method, and small steplike anomalies are observed [10]. In this study, the elastic constants $C_{33}$, $C_{44}$, and $C_{55}$ of UTe$_2$ were precisely measured by the phase comparative method to investigate the lattice instability and the temperature and magnetic field dependences of UTe$_2$ from 2 to 300 K and from 0 to 14 T, respectively.

## 2. Experimental details

A single crystal of UTe$_2$ was grown by the chemical vapor transport method at Tohoku University. We used a single crystal with dimensions of 1.631 × 2.031 × 2.722 mm$^3$ along [100], [010], and [001] directions for this study. A pair of LiNbO$_3$ transducers, which were glued onto the well-polished sample surfaces with room-temperature-vulcanizing silicone rubber, were generated and detected ultrasonic wave. The elastic constant $C_{ij}$ was calculated using the relation $C_{ij} = \rho v_{ij}^2$, where the density $\rho$ is 9.222 g/cm$^3$ at room temperature and $v_{ij}$ is the ultrasonic velocity in the $C_{ij}$ mode. The relations among the elastic constants, the ultrasonic propagation and polarization vectors, strain, and the symmetry (irreducible representation) of the point group $D_{2h}$ for the global symmetry and that of the point group $C_{2v}$ for the local site symmetry of U's 4$i$-site are listed in Table I.

Table I. The relationship between the elastic constants and ultrasonic propagation and polarization vectors, strain, and symmetry in the point group $D_{2h}$ and $C_{2v}$.

| Elastic constant | Propagation vector | Polarization vector | Strain | Symmetry $D_{2h}$ | Symmetry $C_{2v}$ |
|---|---|---|---|---|---|
| $C_{33}$ | [001] | [001] | $\varepsilon_{zz}$ | $\Gamma_1$ ($A_g$) | $\Gamma_1$ ($A_1$) |
| $C_{44}$ | [001] | [010] | $\varepsilon_{yz}$ | $\Gamma_4$ ($B_{3g}$) | $\Gamma_4$ ($B_2$) |
| $C_{55}$ | [001] | [100] | $\varepsilon_{zx}$ | $\Gamma_3$ ($B_{2g}$) | $\Gamma_3$ ($B_1$) |

## 3. Results

Figure 1 shows the temperature dependence of the elastic constants $C_{33}$, $C_{44}$, and $C_{55}$ of UTe$_2$. We use the ultrasonic frequencies of 99,109, and 108 MHz for the $C_{33}$, $C_{44}$, and $C_{55}$ modes, respectively. The elastic constants $C_{33}$ and $C_{44}$ increase with decreasing temperature, whereas the elastic constant $C_{55}$ decreases, that is, elastic softening below ~ 60 K. The elastic constant $C_{33}$ shows a shoulderlike upturn below ~ 20 K, where the thermal expansion coefficient becomes negative. Here, we check whether the negative sign of the thermal expansion coefficient affects the elastic constant $C_{33}$. Considering that the thermal expansion coefficient of UTe$_2$ is $-8.0 \times 10^{-6}$ K$^{-1}$ at 12 K [9], the estimated relative change in length is $\Delta L_c/L_c \sim +1.1 \times 10^{-4}$ in the temperature range from 30 to 2 K, where $L_c$ is the absolute value of the length in the $c$-axis. By using the relationship between the relative change of the sample length (path of the ultrasound) along the $c$-axis and sound velocity $v_{33}$, $\Delta L_c/L_c = \Delta v_{33}/v_{33}$, we can estimate the effect of the length change induced by the negative thermal expansion on the sound velocity change. The relative change in sound velocity at the shoulderlike anomaly around 20 K is $\Delta v_{33}/v_{33} \sim +3.5 \times 10^{-3}$, which is about 30 times larger than the velocity change $\Delta v_{33}/v_{33} \sim -1.1 \times 10^{-4}$ estimated from the length change of the lattice for the $c$-axis. Note that the directions of the length change and the sound velocity change are opposite. Thus, it can be considered that the effect of the length change induced by the negative thermal expansion is negligible and not dominant for the hardening of the elastic constant $C_{33}$. This anomaly obtained from the ultrasonic measurement suggests that the Grüneisen parameter changes with the change in the electronic system, which may be the origin of the negative thermal expansion and also the hardening in the $C_{33}$ mode. Note that, the U-based heavy fermion superconductor UPt$_3$ and dense Kondo compound CeNiSn also show similar elastic anomaly, which is probably responsible for energy gap at the Fermi level in the quasiparticle band [11,12]. The measurements for the other longitudinal elastic constants $C_{11}$ and $C_{22}$ are needed to verify the origin of this behavior in UTe$_2$. For these elastic constants shown in Fig. 1, the most characteristic behavior is the elastic softening in the $C_{55}$ mode. $C_{55}$ takes a local maximum at around 60 K at 0 T, and shows softening of $\Delta C_{55}/C_{55} \sim 9.0 \times 10^{-3}$ with decreasing temperature. At low temperatures below 10 K, the elastic constant $C_{55}$ of UTe$_2$ appears to be leveling off, but the softening continues at least down to 2 K. The softening is observed in only the $C_{55}$ mode, not in the $C_{33}$ and $C_{44}$ modes, at this stage. This marked contrast indicates that the elastic modes are separated properly and strongly suggests that UTe$_2$ has lattice instability in $C_{55}$ related to the symmetrized strain $\varepsilon_{zx}$, with $\Gamma_3$ ($B_{2g}$) symmetry. The present ultrasonic experimental results are not enough for drawing a clear conclusion on the relationship between these lattice instabilities and superconductivity, since we have not made measurements across the superconducting transition point and under high magnetic fields above 35 T. On the other hand, the analogue compound URhGe shows elastic softening in the $C_{66}$ mode whereas no anomalous behavior is observed in the $C_{55}$ [13]. It is proposed that the origin of the elastic softening in URhGe is due to band instability. It is worthwhile to discuss the similarities between these compounds after measuring the transverse $C_{66}$ of UTe$_2$ in order to unravel the possible relationship between superconductivity and lattice instability.

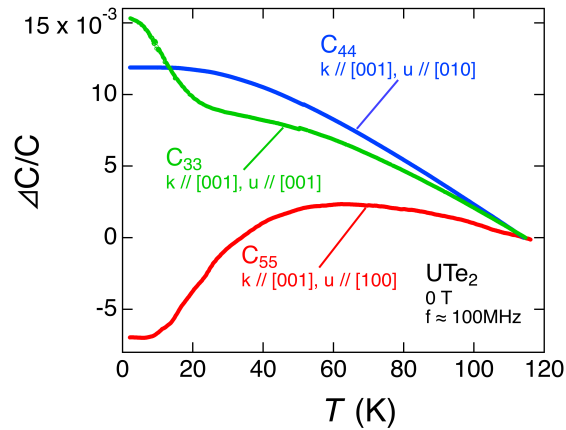

**Fig. 1.** Temperature dependence of the elastic constants $C_{33}$, $C_{44}$, and $C_{55}$ of UTe$_2$ at 0 T. These data are normalized at 115 K. Ultrasonic frequencies are about 100 MHz.

We show the temperature dependence of the elastic constants $C_{33}$, $C_{44}$, and $C_{55}$ under the magnetic field $H \parallel [010]$ in Fig. 2. In the $C_{33}$ mode (Fig. 2(a)), the temperature at which the shoulderlike anomaly appears shifts lower under the magnetic field $H \parallel [010]$. The hardening of the elastic constant $C_{44}$ saturates at a low temperature at 0 T, but the elastic softening appears below 16 K at the higher magnetic field (Fig. 2(b)). This result in the $C_{44}$ mode suggests the existence of lattice instability in the reentrant superconductivity phase above 15 T (SC2), unlike the low-magnetic-field superconductivity phase below ~ 15 T (SC1) in $H \parallel b$. The elastic constant $C_{55}$ at 0 and that at 3 T almost overlap. However, the local maximum shifts to a lower temperature above 10 T. The elastic softening of $C_{55}$ is insensitive to the magnetic field $H \parallel [010]$, which suggests that the lattice instability remains even in higher magnetic fields.

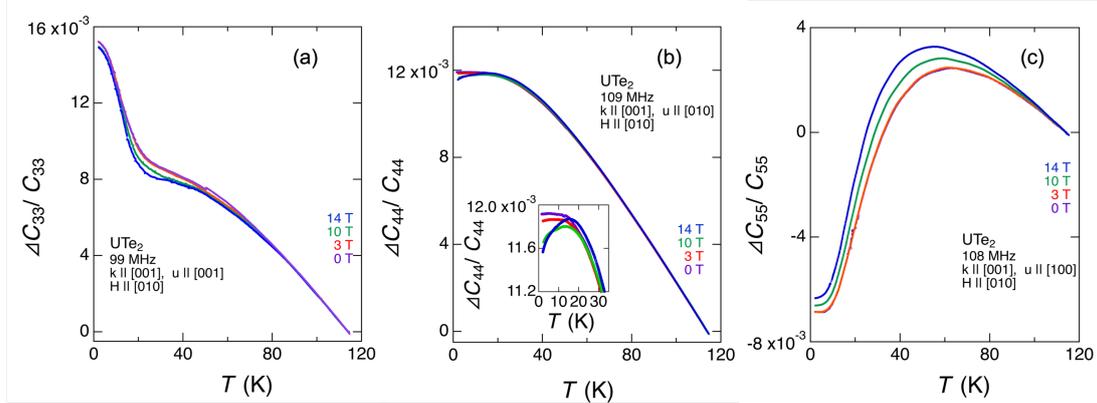

**Fig. 2.** Temperature dependence of the elastic constants (a) $C_{33}$, (b) $C_{44}$, and (c) $C_{55}$ of UTe$_2$ at various magnetic fields for $H \parallel [010]$. These data are normalized at 115 K. Inset (b) is an expanded view of the range of low temperature 0 to 35 K. Ultrasonic frequencies are about 100 MHz. The direction of the magnetic field was adjusted to the eyes.

It has not been clarified whether the 5$f$ electronic state of U in this system can be described by the CEF state. It has also been reported that this system can be explained by the localized 5$f$-electron model above 150 K [14,15]. Further analysis of the softening obtained in this study and its change in a magnetic field using the localized model should help us understand the possible CEF effect of UTe$_2$. As mentioned above, the measurement of another symmetry breaking mode $C_{66}$ for UTe$_2$, which is corresponding to the symmetrized strain $\varepsilon_{xy}$, should also be investigated for comparison and is now in progress, by using a sample shaped with a different geometry.

Finally, we show the magnetic field dependence of the elastic constant $C_{55}$ for $H \parallel$ [010] in Fig. 3. The elastic constant $C_{55}$ increases monotonically with increasing magnetic field at all temperatures. No de Haas–van Alphen oscillation is observed in these temperature and magnetic field ranges. Note that by comparing the magnitude of change in the temperature and magnetic field dependences, the vertical axis of the magnetic field dependence is about 10% of that of the temperature dependence. This result also shows that the elastic softening of $C_{55}$ of UTe$_2$ is insensitive to the magnetic field.

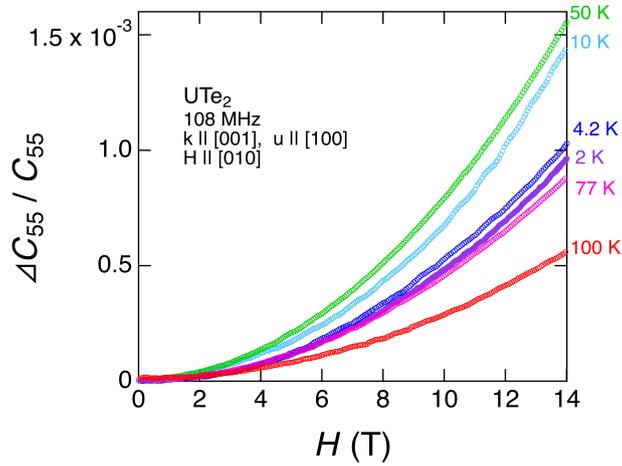

**Fig. 3.** Magnetic field dependence of the elastic constants $C_{55}$ of UTe$_2$ at fixed temperatures for $H \parallel$ [010].

## 4. Conclusion

We have measured the elastic constants of a single-crystal UTe$_2$ sample by ultrasonic measurements as a function of temperature and magnetic field. Relatively large elastic softening is observed only in the $C_{55}$ mode, whereas the $C_{33}$ and $C_{44}$ modes show hardening down to 2 K. The softening of $C_{55}$ remains even under the magnetic field of up to 14 T for $H \parallel$ [010]. To investigate the possible connection between the elastic anomalies found in this work and the presence of SC2 at low temperatures, measurements under a higher magnetic field ($H > 14$ T) and low temperatures are needed. The relationship between the lattice instability of the $C_{55}$ mode, which appeared in the normal state, and SC1 of UTe$_2$ is also unclear. Thus, the measurements below 2 K under low magnetic field are also required. Such ultrasonic measurements of UTe$_2$ in higher magnetic fields and at lower temperatures in the superconducting phase are now in progress. We also plan to continue experiments on the $C_{11}$, $C_{22}$, and $C_{66}$ modes, which could not be measured in this study owing to sample geometry issues.

**Acknowledgment**

This work was supported by JSPS KAKENHI Grant Numbers JP21KK0046, JP22K03501, JP22H04933, JP19H00646, and JP22H04933. R. H. was supported by the JSPS Overseas Challenge Program for Young Researchers. This study was partly supported by Hokkaido University, Global Facility Center (GFC), Advanced Physical Property Open Unit (APPOU), funded by MEXT under the Support Program for Implementation of New Equipment Sharing System Grant No. JPMXS0420100318.